\documentclass[aps,prx,twocolumn,superscriptaddress,showkeys,10pt]{revtex4-2}

\usepackage[utf8]{inputenc}
\usepackage{graphicx}
\usepackage{amsmath}
\usepackage{amssymb}
\usepackage{array}
\usepackage{multirow}
\usepackage{float}
\usepackage{color}
\usepackage{ulem}
\usepackage[T1]{fontenc}
\usepackage{hyperref}
\hypersetup{breaklinks=true,hidelinks,colorlinks=true,urlcolor=blue,citecolor=blue,linkcolor=blue}
\def\vq{{\bf q}}
\def\vk{{\bf k}}
\def\vr{{\bf r}}

\begin{document}

\title{Impact of electron correlations on two-particle charge response in electron- and hole-doped cuprates}

\author{Abhishek Nag}
\email[Email: ]{abhishek.nag@ph.iitr.ac.in}
\affiliation{Diamond Light Source, Harwell Campus, Didcot OX11 0DE, United Kingdom}
\affiliation{SwissFEL, Paul Scherrer Institute, 5232, Villigen-PSI, Switzerland}
\affiliation{Department of Physics, Indian Institute of Technology Roorkee, Uttarakhand 247667, India}
\author{Luciano Zinni}
\affiliation{Facultad de Ciencias Exactas, Ingenier\'{i}a y Agrimensura (UNR), Avenida Pellegrini 250, 2000, Rosario, Argentina}
\author{Jaewon Choi}
\affiliation{Diamond Light Source, Harwell Campus, Didcot OX11 0DE, United Kingdom}
\affiliation{Department of Physics, Korea Advanced Institute of Science and Technology (KAIST), 291 Daehak-ro, Daejeon 34141, Republic of Korea}
\author{J. Li}
\affiliation{Diamond Light Source, Harwell Campus, Didcot OX11 0DE, United Kingdom}
\affiliation{National Synchrotron Light Source II, Brookhaven National Laboratory, Upton, New York 11973, USA}
\author{Sijia Tu}
\affiliation{Beijing National Laboratory for Condensed Matter Physics, Institute of Physics, Chinese Academy of Sciences, Beijing 100190, China}
\affiliation{University of Chinese Academy of Sciences, Beijing 100049, China}
\author{A. C. Walters}
\affiliation{Diamond Light Source, Harwell Campus, Didcot OX11 0DE, United Kingdom}
\author{S. Agrestini}
\affiliation{Diamond Light Source, Harwell Campus, Didcot OX11 0DE, United Kingdom}
\author{S. M. Hayden}
\affiliation{H. H. Wills Physics Laboratory, University of Bristol, Bristol BS8 1TL, United Kingdom}
\author{Mat\'{\i}as Bejas}
\affiliation{Facultad de Ciencias Exactas, Ingenier\'{i}a y Agrimensura and Instituto de Física Rosario (UNR-CONICET), Avenida Pellegrini 250, 2000, Rosario, Argentina}
\author{Zefeng Lin}
\affiliation{Beijing National Laboratory for Condensed Matter Physics, Institute of Physics, Chinese Academy of Sciences, Beijing 100190, China}
\affiliation{University of Chinese Academy of Sciences, Beijing 100049, China}
\author{H. Yamase}
\affiliation{Research Center for Materials Nanoarchitectonics (MANA), National Institute for Materials Science (NIMS), Tsukuba 305-0047, Japan}
\author{Kui Jin}
\affiliation{Beijing National Laboratory for Condensed Matter Physics, Institute of Physics, Chinese Academy of Sciences, Beijing 100190, China}
\affiliation{University of Chinese Academy of Sciences, Beijing 100049, China}
\author{M. Garc\'{\i}a-Fern\'{a}ndez}
\affiliation{Diamond Light Source, Harwell Campus, Didcot OX11 0DE, United Kingdom}
\author{J. Fink}
\email[Email: ]{j.fink@ifw-dresden.de}
\affiliation{Leibniz Institute for Solid State and Materials Research Dresden, Helmholtzstr. 20, D-01069 Dresden, Germany}
\affiliation{Institut f\"ur Festk\"orperphysik, Technische Universit\"at Dresden, D-01062 Dresden, Germany}
\author{Andr\'{e}s Greco}
\email[Email: ]{agreco@fceia.unr.edu.ar}
\affiliation{Facultad de Ciencias Exactas, Ingenier\'{i}a y Agrimensura and Instituto de Física Rosario (UNR-CONICET), Avenida Pellegrini 250, 2000, Rosario, Argentina}
\author{Ke-Jin Zhou}
\email[Email: ]{kejin.zhou@diamond.ac.uk}
\affiliation{Diamond Light Source, Harwell Campus, Didcot OX11 0DE, United Kingdom}

\date{\today}

\begin{abstract}
Estimating many-body effects that deviate from an independent particle approach, has long been a key research interest in condensed matter physics. Layered cuprates are prototypical systems, where electron-electron interactions are found to strongly affect the dynamics of single-particle excitations.  It is however, still unclear how the electron correlations influence  charge excitations, such as plasmons, which have been variously treated with either weak or strong correlation models. In this work, we demonstrate the hybridized nature of collective valence charge fluctuations leading to dispersing acoustic-like plasmons in hole-doped La$_{1.84}$Sr$_{0.16}$CuO$_{4}$ and electron-doped La$_{1.84}$Ce$_{0.16}$CuO$_{4}$ using the two-particle probe, resonant inelastic x-ray scattering. We then describe the plasmon dispersions in both systems, within both the weak-coupling mean-field Random Phase Approximation (RPA) and strong-coupling \textit{t}-\textit{J}-\textit{V} model in a large-$N$ scheme. The \textit{t}-\textit{J}-\textit{V} model, which includes the correlation effects implicitly, accurately describes the plasmon dispersions as resonant excitations outside the single-particle intra-band continuum. In comparison, a quantitative description of the plasmon dispersion in the RPA approach is obtained only upon explicit consideration of re-normalized electronic band parameters. Our comparative analysis shows that electron correlations significantly impact the low-energy plasmon excitations across the cuprate doping phase diagram, even at long wavelengths. Thus, complementary information on the evolution of electron correlations, influenced by the rich electronic phases in condensed matter systems, can be extracted through the study of two-particle charge response.

\end{abstract}


 
\maketitle
\section{\label{sec:Intro}Introduction}
Interactions among constituent entities leading to emergent phenomena is observed across disciplines, including superconductivity~\cite{proust_remarkable_2019,keimer_quantum_2015}, active colloids~\cite{kaiser_flocking_2017}, and neural functions~\cite{chialvo_emergent_2010}. In condensed matter systems, dynamic behavior of the constituent entities is probed using spectroscopic techniques.  For instance, in many-electron systems where electron-electron interactions dominate the low-energy physical properties, angle-resolved photoemission (ARPES) or tunneling spectroscopy can assess the strength of `electron correlation'. These correlation effects arise from short-range  interactions between particles, which are seen in the low energy quasi-particle properties. The direct observation of dynamical charge susceptibility, representing the two-particle charge-charge correlation function $\chi_c''(\vq,\omega)$, in comparison, is possible via spectroscopic techniques such as Resonant Inelastic X-ray Scattering (RIXS) or Electron Energy-Loss Spectroscopy (EELS).

One of the fundamental bosonic excitations in metallic systems, is plasmon, originating from collective charge density oscillations in the presence of long-range Coulomb interactions~\cite{pines_collective_1952}. Typically, in isotropic electron systems, the long-wavelength plasmon energy associated with the charge oscillations is finite. In layered-3-dimensional (3D) electron systems, the 3D Coulomb interactions are poorly screened due to the confinement of charges to planes separated by dielectric blocks. Thus, although 3D Coulomb interactions tend to forbid gapless plasmons~\cite{bozovic_plasmons_1990}, for particular momenta perpendicular to the layers ($q_{z}$), charge oscillations that are out-of-phase may lead to formation of acoustic plasmons ($\omega \to 0$ as $q \to 0$) along with the gapped (optical) plasmons~\cite{fetter_electrodynamics_1974,apostol_plasma_1975,markiewicz_acoustic_2008,grecu_plasma_1973}. 

In layered systems like cuprates, optical plasmons were detected soon after the discovery of the high-$T_C$ superconductivity using Transmission-EELS (T-EELS)~\cite{nucker_plasmons_1989}. Acoustic-like plasmons in the cuprates, however, have only recently been observed with the development of the RIXS technique~\cite{hepting_three-dimensional_2018,lin_doping_2020,singh_acoustic_2022,hepting_evolution_2023,li_charge_2023,bejas_plasmon_2024}. Due to the low-energy of acoustic plasmons, their role has been discussed sparsely since the discovery of superconductivity~\cite{kresin_layer_1988,levallois_temperature-dependent_2016,bill_electronic_2003,bauer_impact_2009,falter_nonadiabatic_1994}.  More importantly, the cuprate superconductors, exhibiting anomalous electronic properties such as those observed in the pseudogap and the strange-metal phases~\cite{keimer_quantum_2015,proust_remarkable_2019}, are widely studied for correlated electron physics. The observation of long wavelength low-energy quantum fluctuations of charges along with spins~\cite{dellea_spin_2017,meyers_doping_2017,hepting_evolution_2023,hepting_three-dimensional_2018,robarts_anisotropic_2019} has, therefore, renewed efforts to develop a unified understanding of electron correlations, Coulomb and exchange interactions aiming towards a microscopic theory~\cite{spalek_superconductivity_2022,fidrysiak_unified_2021,greco_plasmon_2016,greco_origin_2019}.

Correlation effects in cuprates lead to an enhancement of the quasiparticle electron mass.  The mass enhancement factor can be denoted as $m^{\ast}/m$, where $m^{\ast}$ and $m$ are the band mass in the presence of interactions and that predicted by tight-binding calculations, respectively~\cite{comin_arpes:_2015}. In contrast, a variety of descriptions can be found for the plasmon excitations in the cuprates. The dispersion of the optical plasmons observed using T-EELS has been described within mean-field Random Phase Approximation (RPA) theories without explicitly considering electron correlations~\cite{Nuecker_plasmon_1991}. The dispersion of the acoustic-like plasmons observed using RIXS has been described using both free-electron layered models and models incorporating strong electron correlations such as the $t$-$J$-$V$ model ~\cite{lin_doping_2020,hepting_three-dimensional_2018,hepting_evolution_2023,greco_origin_2019,nag_detection_2020,singh_acoustic_2022,greco_plasmon_2016}. The charge carrier doping dependence of plasmon energies could not be explained within an RPA model, leading to the introduction of a  scaling factor to the plasmon energies~\cite{lin_doping_2020,hepting_three-dimensional_2018}. Recently, the $t$-$J$-$V$ model in a large-$N$ approximation was employed to explain the low-doping range dependence of plasmons\cite{hepting_evolution_2023,greco_origin_2019}. A similar doping dependence was also discussed in~\cite{prelovsek_electron-energy_1999}, but only for the optical plasma frequency. In the strange metal phase, momentum-independent broad continua observed using Reflection-EELS~\cite{mitrano_anomalous_2018,husain_crossover_2019} have been described using holographic theories~\cite{romero-bermudez_anomalous_2019}, while a RIXS study has found dispersive excitations in this phase~\cite{hepting_evolution_2023}. This multitude of descriptions raises a pertinent question: Do electron correlations that affect the single-particle excitations strongly, have any role to play in the collective charge excitations in cuprates, and what should be the appropriate framework used to describe it?\\

\begin{figure*}
    \centering
    \includegraphics[width=\linewidth]{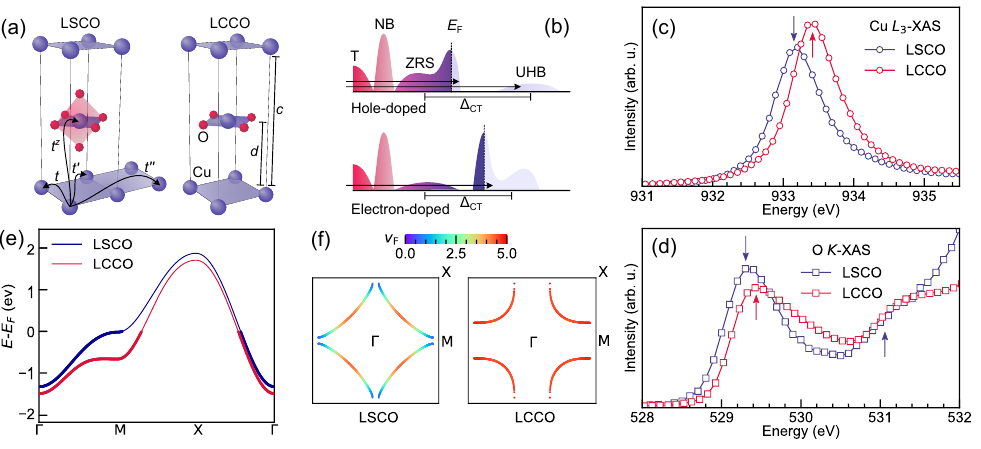}
    \caption{\textbf{Electronic structure of doped cuprates}. (a) Schematic lattice structures of single-layered hole-doped LSCO and electron-doped LCCO showing the Cu-O planes and hopping pathways. O atoms are shown only around the central Cu atom. (b) Schematic representation of electron transitions within the multi-band structure of hole- and electron-doped cuprates~\cite{moritz_effect_2009}. UHB, ZRS, NB and T represent the Upper Hubbard, Zhang-Rice Singlet, Non-Bonding O, and the Zhang-Rice Triplet bands, respectively. (c) Cu $L_3$-edge XAS of LSCO and LCCO. (d) O $K$-edge XAS of LSCO and LCCO ($\delta=0.16$). Arrows mark the photon energies used to probe the plasmons. (e) One-band tight-binding electron dispersion of LSCO and LCCO (see Eq.~\ref{Ek} in Sec.~\ref{subsec:Meth_Theo}). (f) Fermi velocity distribution in LSCO and LCCO.}
    \label{fig:Intro}
\end{figure*}
In this study, we provide a unified perspective on the importance of electron correlations on the dispersion of the acoustic-like plasmons in electron- and hole-doped cuprates probed by RIXS. We compare equal doping levels ($\delta=0.16$) of archetypal hole-doped La$_{1.84}$Sr$_{0.16}$CuO$_{4}$ (LSCO), and  electron-doped La$_{1.84}$Ce$_{0.16}$CuO$_{4}$ (LCCO). The similar lattice parameters of these systems enable investigation of the plasmons in the same momentum phase space. We observe dispersive coherent excitations for both O $K$- and Cu $L_3$-edge RIXS in both systems. For the equal doping level, we find that plasmon velocity in LSCO is smaller than that of LCCO, consistent with the former's smaller Fermi velocity derived from bare band electronic dispersion. However, within a free electron model, the plasmon velocities are overestimated when considering the bare Fermi velocities for both systems. We demonstrate that an appropriate fit to experimentally observed plasmon dispersion is possible within an RPA model with the inclusion of a system dependent band renormalization parameter, and without which, unrealistic values of dielectric constants and incoherent excitations are obtained. The acoustic-like plasmons can be accurately described by the $t$-$J$-$V$ model, where bare band parameters provided as input get implicitly renormalized by electron correlations. Our findings reveal that plasmon dispersion in cuprates is affected by electron correlations like the single-particle excitations, and is accounted for by the  band renormalization parameter in the RPA model. Thus by comparing plasmon dispersions and bare band electron dispersion parameters, it is possible to assess the role and magnitude of electron correlations in different phases in the cuprates.

\section{\label{sec:Res}Results}
\subsection{\label{subsec:XAS}Electronic structure of LSCO and LCCO}
Hole-doped LSCO and electron-doped LCCO belong to the family of single-layered cuprates, obtained upon doping parent systems La$_{2}$CuO$_4$.  They crystallize in distinct structures, the K$_2$NiF$_4$-type $T$ (LSCO) and Nd$_{2}$CuO$_4$-type $T'$ (LCCO)~\cite{das_electronic_2009}. In the $T$ structure, O atoms form octahedral cages around Cu, while apical O atoms are absent in the Cu-O planes in the $T'$ structure [see Fig.~\ref{fig:Intro}(a)], leading to different electronic ground states for the doped systems. A strong Cu-O hybridization and on-site Coulomb interactions give rise to the Upper Hubbard band (UHB) and the Zhang-Rice singlet (ZRS) band in cuprates, as shown in Fig.~\ref{fig:Intro}(b)~\cite{moritz_effect_2009,weber_apical_2010,lee_doping_2006,zaanen_band_1985,emery_theory_1987,varma_charge_1987,chen_electronic_1991,armitage_progress_2010,fink_electron_1994}. The electrostatic potential at the Cu sites is raised due to the lack of apical oxygen in LCCO compared to LSCO, resulting in a reduced charge transfer energy ($\Delta_{\mathrm{CT}}$). Hole-doping shifts the chemical potential ($\mu$) to the ZRS, whereas electron-doping shifts it to the bottom of the UHB. The charge carrier dynamics in these systems can therefore be investigated using x-ray spectroscopy by tuning the photon energy to resonant transitions to these bands. The X-ray Absorption Spectra (XAS) of LSCO and LCCO ($\delta=0.16$), obtained at the Cu $L_3$- and O $K$-edge, respectively, are shown in Fig.~\ref{fig:Intro}(c) and (d).
The Cu $L_3$-edge XAS peak corresponds to transition to the UHB in both systems. In LSCO, the first peak in O $K$-edge XAS corresponds to the hole-states, with the transition to the UHB occuring $1.5$ eV higher~\cite{chen_electronic_1991,fink_electron_1994,pellegrin_orbital_1993}. In LCCO, the first peak in the O $K$-edge XAS is the transition to the UHB, lowered in energy due to the reduced $\Delta_{\mathrm{CT}}$ and chemical shift of the O $1s$ level, consistent with observations in electron-doped Nd$_{2-x}$Ce$_{x}$CuO$_4$ (NCCO)~\cite{pellegrin_orbital_1993,fink_electron_1994}.

\begin{figure*}
    \centering
    \includegraphics[width=\linewidth]{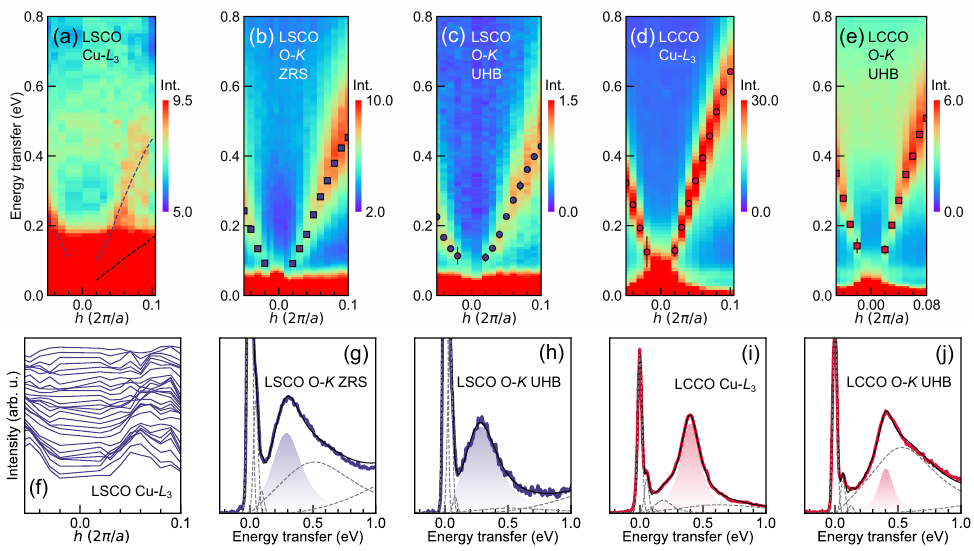}
    \caption{\textbf{Energy-momentum distribution of plasmons in LSCO and LCCO ($\delta=0.16$)}. RIXS intensity maps with incident photon energy at (a) Cu $L_3$-edge, (b) O $K$-edge ZRS and (c) O $K$-edge UHB, respectively, for LSCO. RIXS intensity maps with incident photon energy at (d) Cu $L_3$-edge and (e) O $K$-edge UHB, respectively, for LCCO. For all the data, $k=0.0$ and $l=1.0$. The color scales indicate scattered intensities in arb. u. The markers denote the extracted plasmon energies. In panel (a), the blue dashed line is the plasmon dispersion extracted from panel (b), and the black dashed line is the extended paramagnon dispersion from Ref.~\cite{robarts_anisotropic_2019} for LSCO. Panel (f) shows the momentum distribution curves for energy transfer between 0.225-0.515~eV for Cu $L_3$-edge RIXS on LSCO showing the plasmons. (g)-(j) RIXS line spectra from panels (b)-(e) at $h=0.06$. The dashed lines are elastic, lattice, magnetic and background components as described in Sec. ~\ref{subsec:Meth_Exp}. The shaded distributions are the fitted plasmon peaks which can be compared with the calculated charge susceptibility line profiles in Fig.~\ref{fig:theory}(f-h).}
    \label{fig:RIXS_spectra}
\end{figure*}

\begin{figure}
    \centering
    \includegraphics[width=\linewidth]{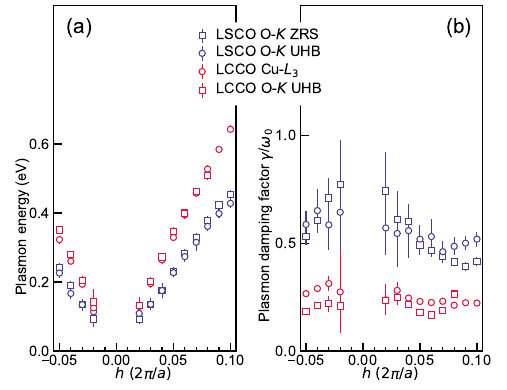}
    \caption{\textbf{Plasmon energies and lifetimes in LSCO and LCCO ($\delta=0.16$)}. (a) Plasmon energies extracted from fits to RIXS spectra at O $K$-edge ZRS and O $K$-edge UHB for LSCO and Cu $L_3$-edge and O $K$-edge UHB for LCCO.  (b) Plasmon damping factor ($\gamma/\omega_0$) extracted from the same fits.}
    \label{fig:RIXS_params}
\end{figure}

\subsection{\label{subsec:RIXS}RIXS}
Fig.~\ref{fig:RIXS_spectra}(a-e) show the RIXS energy-momentum maps collected on LSCO and LCCO ($\delta=0.16$) along the Cu-O in-plane direction $h$, with $k=0.0$ and $l=1.0$. We denote momentum transfers along $h$, $k$, and $l$ directions in reciprocal lattice units, where
$\vq = (ha^{\ast},kb^{\ast},lc^{\ast})$
($a^{\ast}=2\pi/a$, $b^{\ast}=2\pi/b$, $c^{\ast}=2\pi/c$, and $a=b$ and $c$ are the in-plane and out-of-plane lattice parameters, respectively, see Tab.~\ref{tab:LSCO_LCCO}). The incident photon energy for RIXS maps shown in Fig.~\ref{fig:RIXS_spectra}(b) and (c) correspond to resonant transitions to the ZRS and the UHB at O $K$-edge in LSCO, respectively  [see arrows in Fig.~\ref{fig:Intro}(d)]. The highly dispersive plasmon excitations are prominent at both incident energies. A hybridized nature of the doped charges in hole-doped cuprates was concluded based on a similar observation in La$_{1.88}$Sr$_{0.12}$CuO$_{4}$ recently~\cite{singh_acoustic_2022}. Despite the Cu-O hybridized content of the ZRS and the UHB states, study at only the O $K$-edge is insufficient. In Fig.~\ref{fig:RIXS_spectra}(a) we show the RIXS map at Cu $L_3$-edge for LSCO. We can identify faint spectral weight present which appears to follow the plasmon dispersion extracted from O $K$-edge RIXS on LSCO (shown by the blue dashed lines). Presence of these features is also evident in the momentum distribution curves shown in panel (f) for energy transfer between 0.225-0.515~eV. In contrast, our earlier investigation on LSCO and Bi$_2$Sr$_{1.6}$La$_{0.4}$CuO$_{6+\Delta}$~\cite{nag_detection_2020}, along with studies on other hole-doped cuprates~\cite{miao_high-temperature_2017,lee_asymmetry_2014,dellea_spin_2017}, did not reveal the presence of plasmons at Cu $L_3$-edge. This is most likely because the plasmon spectral weight is expected to be the strongest at $l=1.0$~\cite{greco_origin_2019}, where we have investigated in this work. The presence of plasmons at the Cu $L_3$-edge, although extremely weak, validates their hybridized nature in the hole-doped cuprate. In LCCO, the dispersive plasmons are observed clearly for excitation at either Cu $L_3$- or O $K$-edge, as shown in the RIXS maps in Fig.~\ref{fig:RIXS_spectra}(d) and (e), respectively.

Representative fits to the plasmon excitations in the RIXS line profiles as described in Sec.~\ref{subsec:Meth_Exp}, are shown in Fig.~\ref{fig:RIXS_spectra}(g-j). The plasmon energies and widths extracted from the fits are presented in Fig.~\ref{fig:RIXS_params}. It is clear from the similarity of plasmon energies and the widths, that we probe the same charge oscillations at Cu $L_3$ and O $K$-edges for LCCO, and the ZRS and UHB peak at O $K$-edge for LSCO [Fig.~\ref{fig:RIXS_params}(a)]. The plasmons exhibit a nearly linear dispersion for small $h$-values; however, since we cannot resolve the plasmon peaks below $h=0.02$, and a gap may exist at $h,k=0.0$ due to inter-layer hopping ($t_z$)~\cite{hepting_gapped_2022,greco_plasmon_2016}, we describe these excitations to be acoustic-like. Note that an upper limit of $t_z$ was estimated to be 7 meV for LSCO and LCCO~\cite{hepting_gapped_2022,nag_detection_2020}, which is negligibly small to influence the analysis presented in this work. We observe that the plasmon energies for LSCO are smaller than LCCO for the same doping level and at same $h, k=0.0, l=1.0$ values. The plasmons in LSCO are more damped than LCCO as can be seen from Fig.~\ref{fig:RIXS_params}(b) where the extracted damping factor ($\gamma/\omega_0$) is plotted. $\gamma$ and $\omega_0$ represent the plasmon width (damping~$\sim$~inverse lifetime) and plasmon pole energy, respectively. $\gamma/\omega_0$ is found to be less than 1 in the probed momentum phase space, signifying the coherence of the plasmons.

\begin{table*}[t!]
    \caption{Parameters for LSCO and LCCO. 
    In-plane lattice constant: $a$. 
    Distance between the Cu-O layers: $d = c/2$. 
    Doping concentration: $\delta$. 
    Superconducting transition temperature: $T_c$. 
    Average bare Fermi velocity: $\langle v_{\rm F} \rangle^{\rm bare}$. 
    Optical plasmon energy: $\Omega_{\rm p}$. 
    Plasmon velocity obtained using $\langle v_{\rm F} \rangle^{\rm bare}$ in Eq.~\ref{eq:Fett_disp}: $v^{\rm bare}_{\rm p}$. 
    Experimental plasmon velocity: $v^{\rm RIXS}_{\rm p}$. Mass enhancement factor obtained using $v^{\rm RIXS}_{\rm p}$ and renormalized $\langle v_{\rm F} \rangle$ in Eq.~\ref{eq:Fett_disp}: 
    $m^{\ast}/m$.}
    \centering
    \begin{ruledtabular}
    \begin{tabular}{cccccccccccc}
      & $a$ (\AA) & $d=c/2$ (\AA)  & $\delta$ & $T_c$ (K) & $\langle v_{\rm F} \rangle^{\rm bare}$ (eV\AA) & $\Omega_{\rm p}$ (eV)  & $v^{\rm bare}_{\rm p}$ (eV\AA) &$ v^{\rm RIXS}_{\rm p}$  (eV\AA) &  $m^{\ast}/m$ \\
      \hline
LSCO  & 3.77 &6.55  & 0.16 & 38 &2.86 &   0.8~\cite{uchida_optical_1991} &   3.31 & 2.79& 2.0 \\
LCCO  & 4.01 &6.23& 0.16 & 7.87 & 4.58  &    1.2~\cite{lin_doping_2020} &  4.94 & 4.20& 1.7\\ 
    \end{tabular}
    \end{ruledtabular}
    \label{tab:LSCO_LCCO}
\end{table*}

\subsection{\label{subsec:Fett}Effective masses and Fermi velocities in LSCO and LCCO}
\begin{figure*}
    \centering
    \includegraphics[width=\linewidth]{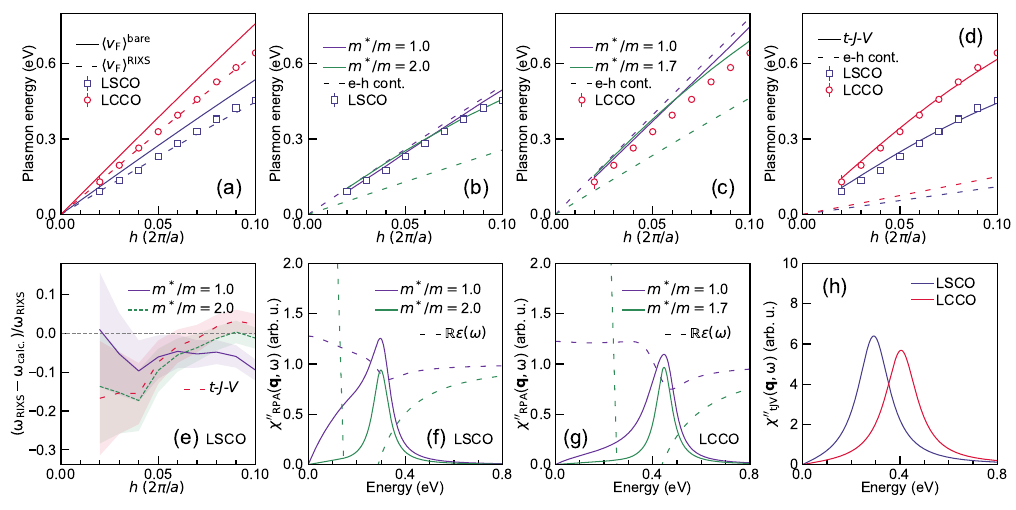}
    \caption{\textbf{Comparison of plasmons to weak- and strong-electron coupling models}. (a) Plasmon energies from RIXS for LSCO and LCCO (markers). Lines are plasmon energies calculated using the free-electron Fetter-Apostol model (Eq.~\ref{eq:Fett_disp}) and different $\langle v_{\rm F} \rangle$s for LSCO and LCCO. (b) Plasmon energies  from RIXS for LSCO (markers). Lines are plasmon energies (continuous) and upper boundaries of electron-hole continua (dashed) calculated using the weak-coupling RPA model with different $m^{\ast}/m$ values for LSCO. (c) Same as in (b) for LCCO. (d) Plasmon energies from RIXS for LSCO and LCCO (markers). Lines are plasmon energies (continuous) and upper boundaries of electron-hole continua (dashed) calculated using the strong-coupling $t$-$J$-$V$ model and bare band parameters for LSCO and LCCO. (e) Momentum dependent deviation of the plasmon energies calculated using the RPA and the $t$-$J$-$V$ models from experiments on LSCO. The shaded areas represent the propagated fitting errors from RIXS spectra. (f) Charge susceptibilities (continuous lines) and  real part of the dielectric functions (dashed lines) at $h=0.06$ obtained from the  RPA model with different $m^{\ast}/m$ values for LSCO. (g) Same as in (f) for LCCO. (h) Charge susceptibilities obtained from the $t$-$J$-$V$ model at $h=0.06$ for LSCO and LCCO. The  charge susceptibility line profiles in panels (f-h) can be compared to corresponding plasmon peaks in RIXS [Fig.~\ref{fig:RIXS_spectra}(g-j)].}
    \label{fig:theory}
\end{figure*}

For the same amount of electron/hole doping, the plasmon energies extracted from RIXS for LSCO are smaller than in LCCO [Fig.~\ref{fig:RIXS_params}(a)], with plasmon velocities $v_{\rm p}^{\rm LSCO}=2.79\pm0.04$~eV\AA~ and $v_{\rm p}^{\rm LCCO}=4.20\pm0.01$~eV\AA~. We first attempt to qualitatively describe the observed plasmon dispersion using the homogenous free-electron layered model, or Fetter-Apostol model (see Eq.~\ref{eq:disfet})~\cite{fetter_electrodynamics_1974,apostol_plasma_1975}. Note that this free-electron model is in the hydrodynamic limit~\cite{fetter_electrodynamics_1974}, or in RPA~\cite{apostol_plasma_1975} and also does not consider interlayer hopping, and as such is not strictly applicable to the cuprates. Although less rigorous compared to many-body models, its simple analytic form allows a rudimentary association of the electronic band parameters to the acoustic plasmon dispersion and the optical plasmon frequency $\Omega_{\rm p}$.  The acoustic plasmon velocity $v_\mathrm{p}$ at $l=1.0$ and small in-plane momentum can be related to the average Fermi velocity $\langle v_{\rm F} \rangle$
using Eq.~\ref{eq:disfet} by,
\begin{equation}
v_\mathrm{p}=\sqrt{\frac{\langle v_\mathrm{F} \rangle^2}{2}+\frac{d^2\Omega^2_\mathrm{p}}{4}} ,
\label{eq:Fett_disp}
\end{equation}

\noindent where, $d$ is the distance between planes.  Assuming that the plasmons are unaffected by electron correlations, we can then use experimentally reported $\Omega_{\rm p}$ (see Tab.~\ref{tab:LSCO_LCCO}) and bare $\langle v_{\rm F} \rangle$ extracted from electron band dispersion to approximately estimate the $v_{\rm p}$. We take tight-binding derived bare parameters for LSCO and NCCO (for LCCO) from Ref.~\cite{markiewicz_one-band_2005} (see Eq.~\ref{Ek} in Sec.~\ref{subsec:Meth_Theo}), and compute the chemical potential $\mu$ for doping $\delta=0.16$. In Fig.~\ref{fig:Intro}(e) we show the bare band dispersion for LSCO and LCCO. The $3d$ band is close to half-filling for hole-doped cuprates while for electron-doped cuprates the band-filling is about 70\%. Due to the proximity to the van Hove filling, this results in a smaller average bare Fermi velocity $\langle v_{\rm F} \rangle^{\rm LSCO, bare}=2.86$  eV{\AA} than  $\langle v_{\rm F} \rangle^{\rm LCCO, bare}=4.58$ eV{\AA} [shown in Fig.~\ref{fig:Intro}(f)]. Using these values in Eq.~\ref{eq:Fett_disp}, we obtain $v_{\rm p}^{\rm LSCO, bare}=3.31$~eV\AA~ and $v_{\rm p}^{\rm LCCO, bare}=4.94$~eV\AA~. It is expected that the plasmon velocities in hole-doped cuprates are smaller than the electron-doped cuprates with similar $d$, due to smaller  $\langle v_{\rm F} \rangle$, however,  as shown in Fig.~\ref{fig:theory}(a), the Fetter-Apostol model with the bare band parameters  overestimates the plasmon velocities by 17\% comparing to the values extracted from RIXS (see Tab.~\ref{tab:LSCO_LCCO}). Conversely, if we use Eq.~\ref{eq:Fett_disp} and $v_{\rm p}$'s extracted from RIXS, we  obtain $\langle v_{\rm F} \rangle^{\rm LSCO} = 1.36$ eV{\AA} and $\langle v_{\rm F} \rangle ^{\rm LCCO} = 2.71$ eV{\AA}. These values are nearly 50\% of the bare band estimates for both systems. Therefore, to explain the experimental results in this approximate model, one needs to use renormalized band dispersions which amount to mass enhancement of $m^{\ast}/m=2.0$ and $m^{\ast}/m=1.7$ for LSCO and LCCO, respectively.

\subsection{\label{subsec:RPA}Random Phase Approximation}
Next, we consider the explicit description of the plasmons within an RPA framework with long-range Coulomb interaction for a layered lattice system (see Sec.~\ref{subsec:Meth_Theo} for details of the implementation). Note that we have ensured that the calculations are consistent with the experimentally reported values of $\Omega_{\rm p}$ for both systems. In Fig.~\ref{fig:theory}(b) and (c), we show the plasmon dispersion extracted from plasmon peaks in $\chi''_{\rm RPA}(\vq,\omega)$ calculated with bare and renormalized band parameters so that $m^{\ast}/m=1.0$ and $m^{\ast}/m=2.0$ for LSCO, and $m^{\ast}/m=1.0$ and $m^{\ast}/m=1.7$ for LCCO. Also plotted, are the upper boundaries of the electron-hole continua for the respective $m^{\ast}/m$ values. For both systems, in the long-wavelength limit, the agreement with the experimental results appears to be slightly better for $m^{\ast}/m=1.0$, while above $h=0.06$, the calculated results for $m^{\ast}/m>1.0$ have smaller deviations from experiments. The momentum dependent deviation from the experimental plasmon energies for LSCO is highlighted in panel (e), showing the better agreement with $m^{\ast}/m>1.0$ for larger momenta and energies. It should also be noted from panels (b) and (c) that, for $m^{\ast}/m=1.0$, the plasmons are within the continuum boundary, while for $m^{\ast}/m>1.0$, they are clearly above the continuum. Panels (f) and (g) show the $\chi''_{\rm RPA}(\vq,\omega)$ and the real part of the dielectric function $\mathbb{R}\epsilon(\omega)$ for $h=0.06$, which undergoes a sign change only for $m^{\ast}/m>1.0$. This signifies that true plasmon resonances which are long-lived are obtained only for $m^{\ast}/m>1.0$. We can compare this observation to the experimentally extracted ratio $\gamma/\omega_0$ [see Fig.~\ref{fig:RIXS_params}(b)]. The $\gamma/\omega_0$ values are less than 1, which mean that experimentally we observe the plasmons as coherently propagating excitations. Additionally, the ratio of in-plane to out-of-plane dielectric constants obtained from the RPA analysis (see Tab.~\ref{tab:theo_tab} in Sec.~\ref{subsec:Meth_Theo}), for $m^{\ast}/m>1.0$ is 1.22 for LSCO and 1.32 for LCCO,
while for $m^{\ast}/m=1.0$ the respective ratios are unrealistic ($\ll1$): 1/6 and 1/4. Thus, the layered lattice RPA model also suggests the use of renormalized band parameters for both systems for describing the plasmons.

\subsection{\label{subsec:tjv}\textit{t}-\textit{J}-\textit{V} model}
In this section, we model the observed plasmon dispersion with the $t$-$J$-$V$ model with long-range Coulomb interaction in a large-$N$ approximation for a layered lattice system, where our inputs are the bare band parameters (see Sec.~\ref{subsec:Meth_Theo} for details of the implementation). Once again, we have verified that the calculations are consistent with the experimentally reported values of $\Omega_{\rm p}$ for both systems. In Fig.~\ref{fig:theory}(d), we show that there is a good agreement between the plasmon dispersions obtained experimentally and those extracted from plasmon peaks in the calculated $\chi''_{\rm tJV}(\vq,\omega)$. The plasmons appear as well-defined peaks [Fig.~\ref{fig:theory}(h)]
and above the electron-hole continuum. This is because the bare band parameters are implicitly renormalized by electron correlations within the theory. To have an estimation of the band renormalization one can see Eq.~\ref{eq:effpar} in Sec.~\ref{subsubsec:tjv_rpa}, which gives $m^{\ast}/m$ of around 4.5 for both systems. Also, the ratio of the in-plane to out-of-plane dielectric constants obtained from the $t$-$J$-$V$ analysis is found to be 1.35 for LSCO and 1.46 for LCCO (see Tab.~\ref{tab:theo_tab} in Sec.~\ref{subsec:Meth_Theo}). Thus, the strongly correlated electron model also describes the plasmons appropriately, without explicitly invoking renormalized band parameters.
 
\section{\label{sec:Disc}Discussion}
\subsection{\label{subsec:dispersion_discussion}Plasmon dispersion and correlations}
Despite the large diversity in material dependent properties, the correlated electron nature of cuprates is widely acknowledged. While the single-particle electron excitations in cuprates clearly show the effects of correlations like mass enhancement and incoherence, charge excitation like plasmons have been described using theories ranging from free-electron to weak- and strong-coupling. The optical plasmon energy $\Omega_{\rm p}$ for zero momentum in the mean-field RPA of homogeneous layered electron systems is proportional to $\sqrt{1/m^{\ast}}$~\cite{fink_electronic_2001,grigoryan_determination_1999}. The optical plasmon dispersion up to second order in $q$ in this model is $\Omega_{\rm p}+Aq^2$, where $A$ is a dispersion coefficient dependent on $m^{\ast}$. Even so, the optical plasmon dispersion observed in Bi$_2$Sr$_2$CaCu$_2$O$_8$ using T-EELS could be described using the bare band parameters~\cite{fink_electronic_2001,grigoryan_determination_1999}. Notably, in Sr$_2$RuO$_4$, a system for which ARPES estimated $m^{\ast}/m \approx 4$, optical plasmons observed using T-EELS  have been modeled using bare band parameters~\cite{knupfer_propagating_2022,schultz_optical_2024}. This was explained on the basis of optical spectroscopy data~\cite{stricker_optical_2014} which found an energy dependent $m^{\ast}/m$: close to 4 below 0.2 eV and close to 1 at higher energies. Resilient quasiparticles at high energies have been predicted by DFT+DMFT calculations~\cite{Deng2013}, and it appears that in Sr$_2$RuO$_4$ the high energy plasmons of 1.5 eV, are unaffected by correlations.

Our observations extend this discussion by focusing on the low-energy acoustic-like plasmon dispersion in LSCO and LCCO, in which the situation seems to be different from the aforementioned. We observe that the plasmon velocity of LSCO is approximately 1.5 times smaller than LCCO for a doping $\delta=0.16$. Since the plasmons are collective excitations involving electrons near the Fermi surface,  one can qualitatively explain this observation by considering the 1.5 times smaller $\langle v_{\rm F} \rangle$ in the hole-doped cuprate. However, when using the $\langle v_{\rm F} \rangle$s derived from bare bands, the plasmon energies are overestimated for both systems in the free electron model. Thus, for cuprates it seems that the acoustic-like plasmons can not be described using bare band parameters and it is necessary to consider the effects of correlation for a quantitative analysis. Within the RPA approach, the agreement of the dispersion with $m^{\ast}/m=1$ worsens as $q$ increases, while it improves for  $m^{\ast}/m>1$. Although it may seem that the effects of correlation may be fully relaxed in the long-wavelength limit, using the $m^{\ast}/m=1$ band parameters result in plasmons appearing within the electron-hole continuum and unreasonable dielectric constant values for either system in our model. The value of the mass enhancement factor $m^{\ast}/m=2.0$ for LSCO is numerically equal to that measured using ARPES at the nodal point~\cite{Ino_2002}. However, this match should not be overemphasised, given that the result from the plasmons represents an average effect over the entire Brillouin Zone (BZ), which means including the anti-nodal region near the saddle point ($\pi,0 $) with a low $v_{\rm F}$, and the nodal region near  ($\pi,\pi $)/2. Also, smaller $m^{\ast}/m$ values are observed in the RPA models for LCCO than LSCO. Although weaker correlations are expected in electron-doped than in hole-doped cuprates~\cite{weber_strength_2010,kobayashi_electronhole_2003,segawa_zero-doping_2010,mizuno_dmft_2017,ogura_asymmetry_2015,weber_apical_2010,de_medici_correlation_2009,das_electronic_2009}, it should be noted that the value of $m^{\ast}/m$ for LCCO was obtained using the band parameters of NCCO in the calculations. This is due to unavailability of the band parameters for LCCO.

\begin{figure}
    \centering
    \includegraphics[width=\linewidth]{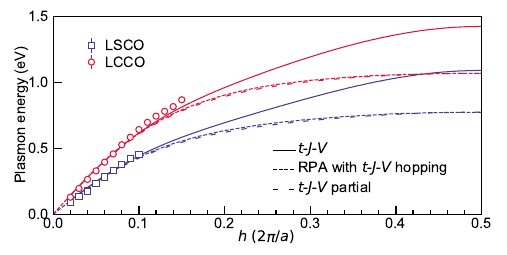}
    \caption{\textbf{Momentum dependence of correlation effects on plasmons.} Plasmon energies from RIXS for LSCO and LCCO (markers). Continuous lines are plasmon energies calculated using the full $t$-$J$-$V$ model. Dense dashed lines are plasmon energies calculated using the RPA model with renormalized band parameters obtained from the $t$-$J$-$V$ model. The match between the two models at long wavelengths suggest that RPA with the renormalized band mass formalism accounts for the effects of electron correlations in this region of momentum space. Locally strong correlation effects stemming from double occupancy prohibition that are absent in RPA, lead to deviations only at large momenta. Sparse dashed lines are plasmon energies calculated using a partial $t$-$J$-$V$ model (see Sec.~\ref{subsubsec:tjv_rpa}). The overlapping dispersions obtained from the RPA and the partial $t$-$J$-$V$ model show that despite the apparent complication of the  $t$-$J$-$V$ formalism with respect to RPA, it has a "hidden" RPA structure including the effects of the electronic correlations.}
    \label{fig:RPA_tJV}
\end{figure}

It can be seen from Fig.~\ref{fig:theory}(e) that the deviation from the experiments in the $t$-$J$-$V$ model, in which the correlation effects are implicit, is similar to that obtained from RPA for $m^{\ast}/m>1$ [smaller (larger) difference at high (low) $q$]. In the $t$-$J$-$V$ model, the bare band parameters get renormalized by the doping $\delta$ and $J$, and additionally the charge response contains fluctuations of the constraint that prohibits double occupancy at a given site. To compare with the RPA model, we use the renormalized band parameters obtained from $t$-$J$-$V$ in the RPA and plot the calculated plasmon dispersions for LSCO and LCCO in Fig.~\ref{fig:RPA_tJV}. We observe that in the long-wavelength region, the $t$-$J$-$V$ and RPA plasmon dispersions coincide. However, at short-wavelengths, the RPA plasmon dispersions deviate from the $t$-$J$-$V$. In  Fig.~\ref{fig:RPA_tJV}, we also plot plasmon dispersions obtained from the $t$-$J$-$V$ excluding some bosonic self-energy components which carry information of the coupling between charge fluctuations and fluctuations of the Lagrange multiplier that force the non double occupancy constraint (see Sec.~\ref{subsubsec:tjv_rpa}).  Exclusion of these components from the $t$-$J$-$V$ model results in a mathematically identical form of charge susceptibility to RPA and hence identical plasmon dispersions are obtained for the two models. Thus, it is evident that the use of renormalized band parameters, i.e., the inclusion of $m^{\ast}$, in RPA accounts for  electron correlations through the enhanced band mass at long wavelengths, while the locally strong correlation effects stemming from double occupancy prohibition that are absent in RPA, lead to deviations only at large momenta. It would be interesting to probe the acoustic-like plasmons till large momenta to: (a) fit the long-wavelength acoustic-like plasmon dispersion using RPA with the inclusion of an effective mass $m^{\ast}$ and using the $t$-$J$-$V$ model, and (b) evaluate the nature of the predicted disagreement between RPA and the $t$-$J$-$V$ model at large momenta ($h > 0.25$).

\begin{figure}
    \centering
    \includegraphics[width=\linewidth]{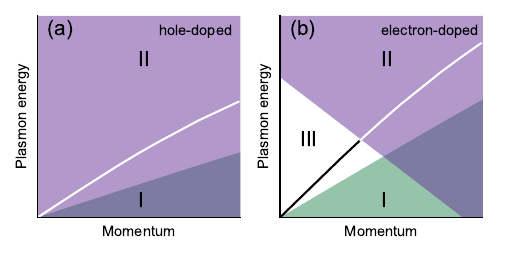}
    \caption{\textbf{Broadening of plasmons due to a decay into single-particle excitation continua}. The electron-hole continuum due to intra-band transitions (I). The continuum due to umklapp scattering related to inter-band transitions (II). Energy-momentum pocket devoid of continuum in electron-doped systems (III). White lines:  Plasmon damping caused by decay into the continuum from single-particle excitations. (a) LSCO, close to half-filled conduction band.  (b) LCCO with additional filling of the conduction band, causing a pocket in the continuum of the inter-band continuum.}
    \label{fig:Band_folding}
\end{figure}

\subsection{\label{subsec:width_discussion}Plasmon Width}
In a free-electron model there is no momentum phase space for the decay of an acoustic or an optical plasmon into intra-band electron-hole excitations (Landau damping) if the plasmon is above the continuum (region I in Fig.~\ref{fig:Band_folding}). Moreover, in the long-wavelength region, the plasmon should be undamped. Nevertheless, non-resolution-limited plasmons are observed in experiments. For finite $q$ a $q^2$ dependence was predicted due to a decay into electron-hole pair excitations, but the theoretical estimates of the broadening were an order of magnitude smaller than the experimentally determined values~\cite{DuBois1969}. Calculations for decay via phonon and impurity assisted intra-band transitions were also found to be insufficient~\cite{Oliveira_1980,Sturm1982}. Thus, the plasmon width had been a puzzle for long time. Finally, it was theoretically proposed~\cite{Paasch_width_1970} that plasmon width appears through a decay into the continuum formed due to inter-band excitations (region II in Fig.~\ref{fig:Band_folding}). The latter originate from \textit{umklapp} processes due to the square of the Fourier transform of the pseudo-potential of the ions in neighbouring BZs. Experimentally, this was supported by EELS on alkali metals, where the plasmon width was found to be proportional to the square of the pseudo-potential~\cite{Gibbons_width_1977,vom_Felde_width_1989}.  
From the $\gamma/\omega_0<1$ values extracted from RIXS [Fig.~\ref{fig:RIXS_params}(b)], we can see that the damping of acoustic plasmons in LSCO is twice as large than LCCO. In the case of a half filled band such as in LSCO and a $k_F$ equal to half of the BZ, the inter-band continuum extends to ($q, \omega=0$). Thus, the acoustic plasmons are damped additionally regardless of the intra-band continuum. Upon changing the band filling, a pocket appears in the inter-band continuum in the low-energy low-momentum region (region III in Fig.~\ref{fig:Band_folding}(b)). In this case (e.g. in LCCO), the acoustic plasmon will be less damped. Here we mention that such pockets causing nearly undamped plasmons at low energy were previously described in T-EELS studies of K-doped graphite~\cite{Ritsko_1981_backfolding}.
In Ref.~\cite{hafermann_collective_2014} the authors considered the Hubbard model in the presence of the long-range Coulomb interaction using dynamical mean field theory, and plasmons were obtained if one-particle self-energy effects and vertex corrections due to correlations are treated properly. The inclusion of electronic self-energy effects leads to a broadening of the plasmons (along with mass enhancement), and an energy dependence of the mass enhancement cannot be ruled out~\cite{stricker_optical_2014,hunter_fate_2023,michon_reconciling_2023}. One- particle self-energy effects can be expected from the interaction between carriers and the rich variety of low-energy charge excitations in the energy scale of $J$~\cite{khaliullin_theory_1996,bejas_dual_2017,zafur_spin_2024}, that may lead to further differences in the plasmon lifetimes of LSCO and LCCO. Further contribution to the broadening of the plasmons may appear from plasmon-phonon decays. Instead of material specific tuning of the broadening, in our calculations we have considered a minimal momentum and energy independent broadening $\Gamma=0.04$ eV (0.1$t$) comparable to experimental energy resolution to study the plasmon behavior. It should be noted that our analysis is performed in the context of the $t$-$J$-$V$ model, i.e., the  $t$-$J$ model, which can be derived from the Hubbard model in the large-$U$ limit, in presence of the long-range Coulomb interaction. Presence of plasmons in our calculations in spite of the strong correlations shows the consistency of our results with that of Ref.~\cite{hafermann_collective_2014}.

\section{\label{sec:conclusion}Conclusions}
We have investigated LSCO and LCCO at equal doping using RIXS, and have observed acoustic-like plasmons having different velocities. We find that the plasmon parameters (energy and lifetime) are identical for a given system irrespective of the probed site (Cu or O). While the RIXS cross-section is typically dominated by local-site effects, this observation highlights that the probed charge excitations are non-local and site-independent, similar to magnetic excitations, due to the hybridized nature of valence charge fluctuations. We show that to appropriately describe the acoustic-like plasmon dispersions in cuprates in a mean-field RPA approach one has to consider renormalized band dispersion parameters. A similar renormalization of the bare band parameters occurs implicitly in the strong-coupling $t$-$J$-$V$ model. This holds true for both sides of the cuprate doping phase diagram, where we observe $m^{\ast}/m>1$ for both LSCO and LCCO. The comparison with the $t$-$J$-$V$ model justifies the use of the renormalized band parameters in the RPA approach to effectively represent the mass enhancement stemming from electron correlations at long wavelengths. Therefore, the weak-coupling nature of the RPA should not be used to dismiss its practical usage in cuprates without due consideration. The role of correlations in the two-particle charge response extends beyond a simple adjustment of band parameters. An enhanced band mass reduces the average Fermi velocity and pushes the electron-hole continuum below the plasmon energies, allowing the observation of plasmons as resonant collective excitations. Even though here we have used a uniform mass enhancement contribution to the acoustic-like plasmon dispersion, optical and ARPES studies on Sr$_2$RuO$_4$ have suggested the mass enhancement factors to be dependent on the quasiparticle energy~\cite{stricker_optical_2014,hunter_fate_2023}. Typically the spin exchange energies ($\approx$ 0.2~eV) are much smaller than the optical plasmon energies of about 1~eV. This may rationalize the non-dependence of optical plasmon dispersion on electron correlations observed using T-EELS in cuprates and ruthenates~\cite{Nuecker_plasmon_1991,knupfer_propagating_2022,schultz_optical_2024,roth_evidence_2020}, in contrast to the acoustic plasmons. The difference in the influence of electron correlations on acoustic and optical plasmons will be the subject of our research in the near future.

\section{\label{sec:Meth}Methods}
\subsection{\label{subsec:Meth_Exp}Experimental details}
Single crystal of La$_{1.84}$Sr$_{0.16}$CuO$_4$ (LSCO) grown by floating-zone method and used for previous report on plasmons~\cite{nag_detection_2020}, was reused for this experiment. The crystal was re-cleaved before measurement at each edge, in vacuum. Hole-doping of $\delta=0.16$ was verified using magnetization measurements of LSCO corresponding to superconducting transition temperature of 38~K.

High-quality La$_{2-x}$Ce$_{x}$CuO$_4$ films were grown on
SrTiO$_3$ substrates via the pulsed laser deposition technique
with 100 nm thickness. The films have a linearly varying Ce concentration ($x=$ 0.1 to 0.19) along the surface of the substrate, fabricated by the continuous moving mask technique~\cite{yuan2022scaling}. The direction of varying concentration is aligned normal to the RIXS scattering plane. The $c$-axis lattice constants and superconducting transition temperatures measured along the concentration gradient direction are consistent with results from single-doping LCCO films~\cite{sawa_electron-doped_2002}. For $x=0.16$, a superconducting transition temperature of 7.87~K was observed using resistivity measurements.

The pressure inside the sample vessel was maintained around $5\times10^{-10}$ mbar. The samples were cooled down to 25~K. While this means that the LSCO was below and the LCCO was above $T_{\rm C}$, a recent article~\cite{hepting_evolution_2023}, did not find significant change in the plasmon dispersion in this temperature range.  The XAS were collected as total electron yield in normal incidence geometry with $\sigma$-polarization, so that the electric field was in the Cu-O plane. High energy-resolution RIXS spectra were collected at Cu $L_3$-($\Delta E \simeq 0.045$ eV) and O $K$- ($\Delta E \simeq 0.043$ eV) edges with $\sigma$-polarization, at the I21-RIXS beamline, Diamond Light Source, United
Kingdom~\cite{zhou_i21:_2022}. The zero-energy transfer position and  energy resolution were determined from subsequent measurements of elastic peaks from an adjacent carbon tape. Negative and positive values of $h$ represent the grazing-incident and grazing-exit geometries, respectively.

RIXS data were normalized to the incident photon flux, and subsequently corrected for self-absorption effects prior to fitting. A Gaussian line-shape with the experimental energy resolution was used to fit the elastic line. Gaussian line-shapes were also used to fit the low energy phonon excitations at $\sim 0.045$~eV and their overtones. The scattering intensities $S$($\vq$, $\omega$) of the plasmons, bimagnons and paramagnons, dependent on the imaginary part of their respective dynamic susceptibilities $\mathbf{\chi}''$($\vq$, $\omega$) were modelled as:
\begin{equation}
S(\vq, \omega) \propto \frac{\mathbf{\chi}''(\vq, \omega)}{1- e^{-\hbar\omega/k_BT} },
\end{equation}
where $k_B$, $T$ and $\hbar$ are the Boltzmann constant, temperature and the reduced Planck constant. A generic damped harmonic oscillator model was used for the response function 

\begin{equation}
\mathbf{\chi}''(\vq, \omega) \propto \frac{\gamma\omega}{\left[\omega^2-\omega^2_0\right]^2+4\omega^2\gamma^2},
\label{dho}
\end{equation}

\noindent where $\omega_0$ and $\gamma$ are the undamped frequency and the damping respectively.

First we extracted the zone-centre energy, amplitude and width of the broad incoherent mode at $h=0.01$ and concluding this to be a bimagnon, fixed its amplitude and width for the whole momentum-range~\cite{nag_detection_2020}. The energy values of the bimagnons were allowed to vary within $\pm20$ meV. An additional paramagnon component was added for the RIXS spectra at Cu $L_3$-edge for LCCO. Significant correlations were found below $h<0.02$, between the elastic, phonon and plasmon amplitudes and energies, and hence the plasmon energy values determined in these regions are less conclusive and not reported.  A high energy quadratic background was also included in the fitting model to account for the tailing contribution from $dd$-excitations above 1.5 eV. Representative fits using this model are shown in Fig.~\ref{fig:RIXS_spectra}(g-j).

\subsection{\label{subsec:Meth_Theo}Theory details}
\subsubsection{Band dispersion and Coulomb repulsion}

Based on \textit{ab-initio} calculations, the electron band dispersion for the cuprates was proposed as: 

\begin{equation}
 E_{\vk} = E_{\vk}^{\parallel}  + E_{\vk}^{\perp} ,
\label{Ek}
\end{equation}

\noindent  where the in-plane dispersion $E_{\vk}^{\parallel}$ (Ref.~\cite{markiewicz_one-band_2005}), and the out-of-plane dispersion 
$E_{\vk}^{\perp}$ are given, respectively, by
\begin{align}
E_{\vk}^{\parallel} =& -2 t(\cos k_{x}+\cos k_{y})
-4t'\cos k_{x} \cos k_{y} \nonumber \\
&-2t''(\cos 2k_{x} + \cos 2k_{y})- \mu  \,, \label{EparaRPA} \\
E_{\vk}^{\perp} =& -\frac{t_{z}}{4} (\cos k_x-\cos k_y)^2 \cos k_z\, \label{EperpRPA} ,
\end{align}
\noindent with  $\mu$ as the chemical potential. The different hopping pathways in the materials are shown in Fig.~\ref{fig:Intro}(a). We use the bare parameters~\cite{markiewicz_one-band_2005}: $t=0.4$~eV, $t'/t=-0.09$, and $t''/t=0.07$ for LSCO, and $t=0.4$~eV, $t'/t=-0.24$, and $t''/t=0.15$ for LCCO. The parameters used for LCCO are those given in Ref.~\cite{markiewicz_one-band_2005} for NCCO, due to unavailability of data for LCCO. In the out-of-plane dispersion $E_{\vk}^{\perp}$ we have replaced $\cos k_z$ by $1$ in the calculation, i.e., the contribution of $E_{\vk}^{\perp}$ is independent of $k_{z}$ which leads to a vanishing plasmon gap at the zone center, even for a finite value of $t_z$. This is justified by the fact that the plasmon gap in LSCO and LCCO, if it exists, is small and at present inaccessible experimentally, a topic which was discussed in depth in Ref.~\cite{hepting_gapped_2022}. In other words the presence of $\cos k_z$, as it was in Refs. \cite{nag_detection_2020,hepting_evolution_2023,hepting_gapped_2022},  and the $E_{\vk}^{\perp}$ dispersion is nearly irrelevant to the present analysis. Without loosing generality, we have assumed $t_z/t=0.01$ for both systems. We have also neglected $t'''$ and $t'_z$. Finally, we compute the chemical potential $\mu$ for each case for doping $\delta=0.16$, which gave $\mu=-0.24$ eV and $\mu=0.038$ eV for LSCO and LCCO, respectively. 

Earlier works~\cite{fetter_electrodynamics_1973,fetter_electrodynamics_1974,grecu_plasma_1973}, considered the long-range Coulomb interaction
$V(\vq)$ for homogeneous layered electron gas as
\begin{equation}\label{eq:lay}
V(\vq)=V(q_{||},q_z) =\frac{q_{||} d }{2} \frac{\sinh(q_{||} d)}{\cosh(q_{||} d)-\cos(q_z d)}.
\end{equation}
Here, we use the long-range Coulomb interaction
$V(\vq)$ for a layered lattice system for the RPA and $t$-$J$-$V$ models: 

\begin{equation}
V(\vq)=\frac{V_c}{A(q_x,q_y) - \cos q_z} ,
\label{LRC}
\end{equation}

\noindent where $V_c= e^2 d(2 \epsilon_{\perp} a^2)^{-1}$ and 

\begin{equation}
A(q_x,q_y)=\alpha (2 - \cos q_x - \cos q_y)+1 .
\end{equation}

These expressions are easily obtained by solving the Poisson's equation on the lattice \cite{becca_charge-density_1996}.  
Here $\alpha=\tilde{\epsilon}/[(a/d)^2]$, $\tilde{\epsilon}=\epsilon_\parallel/\epsilon_\perp$, and $\epsilon_\parallel$ and $\epsilon_\perp$ are the 
dielectric constants parallel and perpendicular to the planes, respectively. It is important to note that in the present $V(\vq)$ model we have two dielectric constants instead of one as in \cite{fetter_electrodynamics_1974,grecu_plasma_1973}.     
$e$ is the electric charge of electrons; 
$a$ is the in-plane lattice constant and the in-plane momentum ${\bf q}_\parallel = (q_x,q_y)$ is calculated 
in units of $a^{-1}$; similarly $d$ is the distance between the Cu-O planes, and the out-of-plane momentum $q_z$ is calculated in units of $d^{-1}$. In the present work, we consider $V_c$ and $\alpha$ as independent parameters, and from them we can estimate $\epsilon_\parallel$ and $\epsilon_\perp$ and discuss their reliability.

\begin{table*}[t!]
    \caption{Parameters extracted by fitting experimental plasmon dispersions to the different models.}
    \centering
    \begin{ruledtabular}
    \begin{tabular}{ccccccc}
      & Model  & $m^{\ast}/m$   & $V_c$ (eV)  & $\alpha$& $\epsilon_\parallel/\epsilon_0$ & $\epsilon_\perp/\epsilon_0$ \\
      \hline
LSCO  & RPA         & 1.0  & 0.49  & 0.5 & 14.1 &  85.1 \\
      & RPA         & 2.0  & 7.6   & 3.7 & 6.72 &  5.49 \\
      & $t$-$J$-$V$ &  -   & 18.8  & 4.1 & 3.01 &  2.22 \\
LCCO  & RPA         & 1.0  & 0.9   & 0.6 & 10.1 & 40.8 \\
      & RPA         & 1.7  & 9.2   & 3.2 & 5.05 &  3.81 \\
      & $t$-$J$-$V$ &  -   & 30.0  & 3.5 & 1.71 &  1.17 \\
    \end{tabular}
    \end{ruledtabular}
    \label{tab:theo_tab}
\end{table*}

\subsubsection{Random phase approximation}

In RPA the charge correlation function is given by the well-known expression~\cite{mahan_many-particle_1990}
\begin{eqnarray}
\chi_{\rm RPA}({\bf q},\mathrm{i}\omega_{n})=\frac{\chi^{(0)}({\bf q},\mathrm{i}\omega_{n})}
{1-V(\vq)\chi^{(0)}({\bf q},\mathrm{i}\omega_{n})} , 
\label{chiRPA}
\end{eqnarray}
\noindent  where $\chi^{(0)}({\bf q},\mathrm{i}\omega_{n})$ is the usual 
Lindhard function, 

\begin{eqnarray}
\chi^{(0)}(\vq,\mathrm{i}\omega_n)= \frac{2}{N_s}\sum_{\vk} 
\frac{n_F(E_{\vk-\vq})-n_F(E_\vk)}
{\mathrm{i}\omega_n-E_\vk+E_{\vk-\vq}},
\label{chi0}
\end{eqnarray}

\noindent which accounts for the particle-hole continuum. $\vq$ is a three dimensional wave-vector, $\omega_n$ is a boson Matsubara frequency, and the factor $2$ comes from the spin summation. $N_s$ is the number of sites in each plane and $n_F$ is the Fermi distribution. The denominator in Eq.~\ref{chiRPA} is the RPA dielectric function $\varepsilon({\bf q},\mathrm{i}\omega_{n})=1-V(\vq)\chi^{(0)}({\bf q},\mathrm{i}\omega_{n})$.

After performing the analytical continuation
$\mathrm{i}\omega_n \rightarrow \omega+\mathrm{i} \Gamma$ in 
$\chi_{\rm RPA}(\vq,\mathrm{i}\omega_n)$, we obtain the imaginary part of the charge-charge correlation functions $\chi''_{\rm RPA}(\vq,\omega)$, which can be directly compared with RIXS. $\Gamma$ influences the width of the plasmon and its effect on the plasmon peak position is strongest when it becomes comparable to undamped plasmon energy (overdamped condition). From Fig.~\ref{fig:RIXS_params}(b), we see that this condition may be applicable only close to the zone-centre. As discussed in the main text, several factors can affect the plasmon width. Here we consider a minimal momentum and energy independent broadening $\Gamma$=0.04 eV (0.1$t$) comparable to experimental energy resolution to study the plasmon behaviour ~\cite{greco_origin_2019,greco_close_2020,prelovsek_electron-energy_1999,hepting_gapped_2022}.
The RPA calculation is a weak coupling approach and, in principle, the electron dispersion $E_{\vk}$ (Eq. \ref{Ek}) is given by the bare band~\cite{markiewicz_one-band_2005}. However, as discussed in the text, the electron hopping parameters $t$, $t'$, and $t''$ are renormalized to account for $m^{\ast}/m>1$. 

The analytical relation between plasmon and Fermi velocities (Eq.~\ref{eq:Fett_disp}) in the homogeneous free-electron layered model by Fetter-Apostol~\cite{fetter_electrodynamics_1974,apostol_plasma_1975} is derived from the plasmon energy: 
\begin{equation}
\omega_\mathrm{p}=\sqrt{\frac{\langle v_{\rm F} \rangle^2}{2} q_\mathrm{\parallel}^2+\Omega_\mathrm{p}^2\frac{q_\mathrm{\parallel} d }{2} \frac{\sinh(q_\mathrm{\parallel} d)}{\cosh(q_\mathrm{\parallel} d)-\cos(q_\mathrm{\perp} d)}},
\label{eq:disfet}
\end{equation}
obtained by using the Coulomb potential in Eq.~\ref{eq:lay} and the denominator in Eq.~\ref{chiRPA}~\cite{fetter_electrodynamics_1974}. Note that in the negligible conduction dissipation limit, the factor $\langle v_{\rm F} \rangle^2 q_\mathrm{\parallel}^2/2$ is absent (for $\omega_\mathrm{p}^2\gg\langle v_{\rm F} \rangle^2 q_\mathrm{\parallel}^2/2$)~\cite{zabolotnykh_plasmons_2020}. This is not the case for the energy-momentum range of plasmons probed in this work.

\subsubsection{\label{subsubsec:tjv}The layered \textit{t}-\textit{J}-\textit{V} model and the large-\textit{N} formalism}

The large-$N$ approach for the $t$-$J$ model was originally developed in Ref.~\cite{foussats_large-$n$_2004}, and extensively used in the context of charge excitations in cuprates, among others, Refs.~\cite{greco_origin_2019,greco_close_2020,nag_detection_2020,hepting_gapped_2022,greco_plasmon_2016,bejas_possible_2012,yamase_electron_2021,yamase_plasmarons_2023}. The aim of this section is to give a brief description of the main formulae. 

The layered $t$-$J$-$V$ model is written as

\begin{align}
H =& -\sum_{i, j,\sigma} t_{i j}\tilde{c}^\dag_{i\sigma}\tilde{c}_{j\sigma} + 
\sum_{\langle i,j \rangle} J_{ij} \left( \vec{S}_i \cdot \vec{S}_j - \frac{1}{4} n_i n_j \right) \nonumber \\
&+ \sum_{\langle i,j \rangle} V_{ij} n_i n_j ,
\label{tJV}  
\end{align}
\noindent where the sites $i$ and $j$ run over a three-dimensional lattice. 
The hopping $t_{i j}$ takes a value $t$, $t'$ and $t''$ between the first, second and third nearest-neighbor 
sites on a square lattice, respectively. The hopping integral between layers is scaled by $t_z$ (see later for the
specific form of the electronic dispersion). 
$\langle i,j \rangle$ denotes a nearest-neighbor pair of sites. 
The exchange interaction $J_{i j}=J$ is considered only inside the plane;  
the exchange term between the planes ($J_\perp$) is much smaller than $J$~\cite{thio_antisymmetric_1988}.
$V_{ij}$ is the long-range Coulomb interaction on the lattice and is given in momentum space by Eq.~\ref{LRC}.
$\tilde{c}^\dag_{i\sigma}$ ($\tilde{c}_{i\sigma}$) is 
the creation (annihilation) operator of electrons with spin $\sigma=(\uparrow, \downarrow)$  
in the Fock space without double occupancy. 
$n_i=\sum_{\sigma} \tilde{c}^\dag_{i\sigma}\tilde{c}_{i\sigma}$ 
is the electron density operator and $\vec{S}_i$ is the spin operator. 

In the large-$N$ theory~\cite{greco_plasmon_2016} the electronic dispersion $E_{\vk}$ reads: 
\begin{equation}
E_{\vk} = E_{\vk}^{\parallel} + E_{\vk}^{\perp} ,
\label{EktJ}
\end{equation}
\noindent where 
\begin{align}
E_{\vk}^{\parallel} =&
-2 \left( t \frac{\delta}{2}+\Delta \right)
                    (\cos k_{x}+\cos k_{y})
-4t' \frac{\delta}{2} \cos k_{x} \cos k_{y} \nonumber \\
&-2t'' \frac{\delta}{2} (\cos 2k_{x} + \cos 2k_{y} )
- \mu  , \label{Epara} \\
E_{\vk}^{\perp} =& -\frac{t_{z}}{4} \frac{\delta}{2} (\cos k_x-\cos k_y)^2 \cos k_z. \label{Eperp}
\end{align}
For a given doping $\delta$, the chemical potential 
$\mu$ and $\Delta$ are determined self-consistently by solving
\begin{equation}{\label {Delta-A}}
\Delta = \frac{J}{4N_s} \sum_{\vk} (\cos k_x + \cos k_y) n_F(E_\vk) , 
\end{equation}
and 
\begin{equation}
(1-\delta)=\frac{2}{N_s} \sum_{\vk} n_F(E_\vk) .
\end{equation}

We have obtained for $\delta=0.16$ the values $\mu=-0.044$ eV and $\Delta=0.024$ eV for LSCO, and $\mu=-0.010$ eV and $\Delta=0.024$ eV for LCCO.

In the context of the $t$-$J$ model using a path-integral representation~\cite{foussats_large-$n$_2004} for Hubbard operators~\cite{hubbard_electron_1963}, a six-component bosonic field is defined as
\begin{equation}
\delta X^{a} = (\delta
R\;,\;\delta{\lambda},\; r^{x},\;r^{y}
,\; A^{x},\;
A^{y}) ,
\label{boson-field}
\end{equation}
where $\delta R$ describes fluctuations of the number of holes at each site, thus, it is related to on-site charge fluctuations, $\delta \lambda$ is the fluctuation of the Lagrange multiplier introduced to enforce the constraint that prohibits the double occupancy at any site, and $r^{x}$ and $r^{y}$ ($A^{x}$ and $A^{y}$) describe fluctuations of the real (imaginary) part of the bond field from the $J$-term. 

The inverse of the $6\times6$ bare bosonic propagator associated with $\delta X^{a}$ is 
\begin{align}\label{D0}
&\left[ D^{(0)}_{ab}({\bf q},\mathrm{i}\omega_{n}) \right]^{-1} = \nonumber \\
&N \left(
 \begin{array}{cccccc}
\frac{\delta^2}{2} \left[ V(\vq)-J(\vq)\right]
& \delta/2 & 0 & 0 & 0 & 0 \\
   \delta/2 & 0 & 0 & 0 & 0 & 0 \\
   0 & 0 & \frac{4}{J}\Delta^{2} & 0 & 0 & 0 \\
   0 & 0 & 0 & \frac{4}{J}\Delta^{2} & 0 & 0 \\
   0 & 0 & 0 & 0 & \frac{4}{J}\Delta^{2} & 0 \\
   0 & 0 & 0 & 0 & 0 & \frac{4}{J}\Delta^{2} \
 \end{array}
\right),
\end{align}

\noindent where   
$J(\vq) = \frac{J}{2} (\cos q_x +  \cos q_y)$. We use $J/t=0.3$. 

At leading order, the bare propagator  $D^{(0)}_{ab}$ is renormalized in
$O(1/N)$. From the Dyson equation the renormalized bosonic propagator is 
\begin{equation}
[D_{ab}(\vq,\mathrm{i}\omega_n)]^{-1}
= [D^{(0)}_{ab}(\vq,\mathrm{i}\omega_n)]^{-1} - \Pi_{ab}(\vq,\mathrm{i}\omega_n) .
\label{dyson}
\end{equation}

Here the $6 \times 6$ boson self-energy matrix $\Pi_{ab}$ is

\begin{align}
\Pi_{ab}(\vq,\mathrm{i}\omega_n)
            =& -\frac{N}{N_s}\sum_{\vk} h_a(\vk,\vq,E_\vk-E_{\vk-\vq}) \nonumber \\ 
  \times &\frac{n_F(E_{\vk-\vq})-n_F(E_\vk)}
    {\mathrm{i}\omega_n-E_\vk+E_{\vk-\vq}} 
            h_b(\vk,\vq,E_\vk-E_{\vk-\vq}) \nonumber \\
        -& \delta_{a\,1} \delta_{b\,1} \frac{N}{N_s}
        \sum_\vk \frac{\tilde{E}_{\vk-\vq}-\tilde{E}_\vk}{2}n_F(E_\vk) ,
        \label{Pi}
\end{align}

\noindent where $\tilde{E}_{\vk}$ is equal to $E_{\vk}$ with $\Delta = 0$ and
the $6$-component interaction vertex is given by 

\begin{align}
 h_a(\vk,\vq,\nu) =& \left\{
                   \frac{2E_{\vk-\vq}+\nu+2\mu}{2} \right. \nonumber\\
                &+ 2 \Delta \left[
                \cos\left(k_x-\frac{q_x}{2}\right)\cos\left(\frac{q_x}{2}\right) \right. \nonumber\\
              &\left. + \cos\left(k_y-\frac{q_y}{2}\right)\cos\left(\frac{q_y}{2}\right)
                \right];1;  \nonumber \\
& -2\Delta \cos\left(k_x-\frac{q_x}{2}\right); -2\Delta \cos\left(k_y-\frac{q_y}{2}\right); \nonumber \\
&\left. 2\Delta \sin\left(k_x-\frac{q_x}{2}\right);  2\Delta \sin\left(k_y-\frac{q_y}{2}\right)
                 \right\} .
\label{vertex-h}
\end{align}

In the writing of this manuscript we noted a misprint in the last term of of Eq.~\ref{Pi} in previous works, which has been corrected here.

As discussed previously~\cite{foussats_large-$n$_2004,bejas_possible_2012}, the element $(1,1)$ of $D_{ab}$ is related to the usual charge-charge correlation function  
$\chi_{tJV} (\vr_i -\vr_j, \tau)=\langle T_\tau n_i(\tau) n_j(0) \rangle$, which in the large-$N$ scheme is  computed in the $\vq$-$\omega$ space as 
\begin{eqnarray}
\chi_{\rm tJV}(\vq,\mathrm{i}\omega_n)= N \left ( \frac{\delta}{2} \right )^{2} D_{11}(\vq,\mathrm{i}\omega_n)  .
\label{CHtJ}
\end{eqnarray}

It is important to remark that the charge-charge correlation function is nearly unaffected by the value of $J$~\cite{bejas_dual_2017}. As for $\chi_{\rm RPA}(\vq,\mathrm{i}\omega_n)$, after performing the analytical continuation
$\mathrm{i}\omega_n \rightarrow \omega+\mathrm{i} \Gamma$ 
in $\chi_{\rm tJV}(\vq,\mathrm{i}\omega_n)$ we obtain the imaginary part of
the charge-charge correlation functions $\chi''_{\rm tJV}(\vq,\omega)$. The plasmon excitations are obtained for the resonant peaks of $\chi''_{\rm tJV}(\vq,\omega)$.

\subsubsection{\label{subsubsec:tjv_rpa}Correlations in the \textit{t}-\textit{J}-\textit{V} and RPA models}
Looking at the large-$N$ formalism in Sec.~\ref{subsubsec:tjv} it is not clear why the plasmon excitations obtained in the context of  the $t$-$J$-$V$ model are similar to those obtained in RPA. Although the large-$N$ formalism for the $t$-$J$-$V$ model seems to be complicated and rather different to the usual RPA, here we show that inside this framework an RPA structure is contained.
The large-{\it N} formalism within the $t$-$J$-$V$ model renormalizes the band parameters due to electron correlations, which can be seen in the band dispersion directly (Eqs. \ref{EktJ}-\ref{Eperp}). Comparing it to the usual tight-binding dispersion (Eqs. \ref{Ek}-\ref{EperpRPA}), we obtain
\begin{equation}
    \begin{split}
    t_{\mathrm{eff}}&=t\delta+\Delta, \\
    t'_{\mathrm{eff}}&=t'\delta, \\
    t''_{\mathrm{eff}}&=t'' \delta, \\ 
    t_{z_{\mathrm{eff}}}&=t_z \delta, 
    \end{split}
    \label{eq:effpar}
\end{equation}
where the hopping parameters $t$, $t'$, $t''$, and $t_z$ are the tight-binding bare ones.  We introduced these effective parameters into the RPA model and plotted the obtained plasmon dispersion in Fig.~\ref{fig:RPA_tJV}. To understand the deviation at large-momenta between RPA and  $t$-$J$-$V$, we consider only the $2\times2$ sector ($a,b=1,2$) in the $D_{ab}\left(\mathbf{q},\mathrm{i}\omega_n\right)$ (Eq.~\ref{dyson}). If in Eq.~\ref{dyson} we set manually the bosonic self-energy components $\Pi_{11}$ and $\Pi_{12}$ to zero, the only relevant component is $\Pi_{22}$, and from Eqs.~\ref{Pi} and~\ref{vertex-h} it can be written as
\begin{align}
    \Pi_{22}\left(\mathbf{q},\mathrm{i}\omega_n\right)=&-N\sum_\mathbf{k} \frac{n_F\left(E_{\mathbf{k}-\mathbf{q}} \right)-n_F\left(E_{\mathbf{k}} \right)}{\mathrm{i}\omega_n - E_{\mathbf{k}}+E_{\mathbf{k}-\mathbf{q}}} \nonumber \\
    =&-N\frac{\chi_0\left(\mathbf{q},\mathrm{i}\omega_n\right)}{2}.
\label{eq:pi_22}
\end{align}
In spite of $\chi_0$ (Eq.~\ref{chi0}) representing the particle-hole continuum within the RPA and $\Pi_{22}$ appearing in the large-$N$ formalism within the $t$-$J$-$V$ model as only one component of the bosonic self-energy carrying the information of the fluctuations of the Lagrange multiplier associated with the constraint that prohibits the double occupancy, both have a similar mathematical form. In this context, we compute $\chi_{\rm tJV}\left(\mathbf{q},\mathrm{i}\omega_n\right)$ in Eq.~\ref{CHtJ} using the physical value $N=2$~\cite{greco_plasmon_2016} which gives
\begin{equation}
    \chi_{\rm tJV}\left(\mathbf{q},\mathrm{i}\omega_n\right)=\frac{\chi_0\left(\mathbf{q},\mathrm{i}\omega_n\right)}{1-V'\left(\mathbf{q}\right)\chi_0\left(\mathbf{q},\mathrm{i}\omega_n\right)}.
\label{eq:chi_c2}
\end{equation}
where $V'\left(\mathbf{q}\right)=2\left[ V\left(\mathbf{q}\right)-J\left(\mathbf{q}\right)\right]$. Equation~\ref{eq:chi_c2} shows that the charge-charge correlation function in the large-$N$ formalism considering only the contribution from $\Pi_{22}$ has an RPA-like mathematical form. This shows the presence of a "hidden"  RPA structure with electronic correlations within the $t$-$J$-$V$ formalism with respect to RPA. The contribution of $J\left(\mathbf{q}\right)$ can be neglected because $V\left(\mathbf{q}\right)$ is significantly larger, and the factor $2$ accounts for the transformation to eV using the renormalized value of $t$ in the large-$N$ formalism. In fact, this supports the picture  that the effective mass $m^{\ast}$ introduced in RPA has an electronic correlated origin.

\begin{acknowledgments}
We thank A. A. Aligia, M. Hepting, C. Falter and W. S. Lee for discussions and comments. The RIXS experiments  were primarily supported by user research program of Diamond Light Source, Ltd., UK through beam-time proposal MM27872. A part of the results presented in this work was obtained by using the facilities of the CCT-Rosario Computational Center, member of the High Performance Computing National System (SNCAD, MincyT-Argentina). J.C. acknowledges financial support from the National Research Foundation of Korea (NRF) funded by the Korean government (MSIT) through Sejong Science Fellowship (Grant No. RS-2023-
00252768). A.N. acknowledges the Marie Sk\l odowska-Curie Grant Agreement No. 884104 (PSI-FELLOW-III-3i). H.Y. was supported by JSPS KAKENHI Grant No. JP20H01856 and World Premier International Research Center Initiative (WPI), MEXT, Japan. Z.L. acknowledges the National Natural Science Foundation of China (12225412),  and CAS Project for Young Scientists in Basic Research (2022YSBR-048). 
\end{acknowledgments}

\end{document}